\documentclass[aps,prx,showpacs,notitlepage,
twocolumn,superscriptaddress,
nofootinbib,preprintnumbers,
floatfix,longbibliography,10pt]{revtex4-1}

\usepackage{tikz-cd}
\usepackage{bbm}
\usepackage{xcolor}
\usepackage{mathrsfs}
\usepackage[commandnameprefix=always]{changes}
\usepackage{epsfig}
\usepackage{url}

\usepackage{balance}

\usepackage{soul,color}
\usepackage{graphicx}
\usepackage{mleftright}
\usepackage{amsfonts}
\usepackage{amsthm}
\usepackage[figuresright]{rotating}
\usepackage{amssymb}
\usepackage{amsmath}
\usepackage{dcolumn}
\usepackage{physics}
\usepackage{float}
\usepackage{bm}
\usepackage{verbatim}
\usepackage{braket}
\usepackage[normalem]{ulem}
\usepackage[ruled,vlined,linesnumbered]{algorithm2e}
\usepackage{setspace}
\usepackage[colorlinks,linkcolor=blue,anchorcolor=blue,citecolor=blue,urlcolor=blue]{hyperref}
\usepackage{stackengine}

\usepackage{tikz}
\usetikzlibrary{quantikz2}

\begin{document}

\title{Analytic Approach to Quantum Control Using Quantum Signal Processing}

\author{Aishwarya Majumdar}
\affiliation{Department of Electrical and Computer Engineering, North Carolina State University, Raleigh, NC 27606, USA}

\author{John M. Martyn}
\affiliation{Physical and Computational Sciences, Pacific Northwest National Laboratory, Richland, WA 99354, USA}
\affiliation{Harvard Quantum Initiative, Harvard University, Cambridge, MA 02138, USA}
\affiliation{Department of Physics, Harvard University, Cambridge, MA 02138, USA}

\author{Yuan Liu}
\email[]{q\_yuanliu@ncsu.edu}
\affiliation{Department of Computer Science, North Carolina State University, Raleigh, NC 27606, USA}
\affiliation{Department of Physics and Astronomy, North Carolina State University, Raleigh, NC 27606, USA}
\affiliation{Department of Electrical and Computer Engineering, North Carolina State University, Raleigh, NC 27606, USA}

\author{Nathan Wiebe}
\email[]{nathan.wiebe@utoronto.ca}
\affiliation{Department of Computer Science, University of Toronto, ON M5S 1A1, Canada}
\affiliation{Physical and Computational Sciences, Pacific Northwest National Laboratory, Richland, WA 99354, USA}

\date{\today}

\begin{abstract}
Realizing coherent quantum computation requires precise and robust manipulation of quantum systems through quantum control protocols. Most quantum control techniques rely on heuristic methods for designing the driving pulses that steer the system towards a target state.
Such methods are often based on brute-force optimization and offer limited understanding of the solution landscape. In contrast, quantum algorithms offer a rich body of analytical methods with rigorous error guarantees for implementing unitary and non-unitary transformations, which suggests a promising direction for developing new approaches to quantum control.
Among various such algorithms, quantum signal processing (QSP) has emerged as a powerful framework for quantum algorithm design, implementation, and optimization. However, its potential for quantum control remains largely unexplored. In this work, we establish \textit{QSP-Control}, an analytical framework for quantum control of qubit-oscillator dynamics. We focus on dispersively coupled qubit-oscillator systems and employ the QSP formalism to mitigate unwanted nonlinear effects arising from cross-Kerr interactions. In addition, we develop constructions for precise manipulation of Fock states by designing Fock-state-selective operators, based on structural parallels between the Jaynes-Cummings interaction and QSP. 
These findings demonstrate how several practically relevant problems in quantum control can be mapped to forms amenable to QSP, offering both a systematic design framework and an interpretable perspective on quantum control. 
\end{abstract}

\maketitle

\tableofcontents

\section{Introduction}

As we advance toward the fault tolerant regime of quantum computing~\cite{preskill1997faulttolerantquantumcomputation, PhysRevA.57.127}, achieving precise control over quantum systems in the presence of adversarial factors such as environmental noise and decoherence \cite{PhysRevA.51.992} becomes increasingly essential to facilitate seamless manipulation, storage, and retrieval of quantum information.
Quantum control methods set the ultimate limit on the achievable fidelity of quantum gate operations  and sit at the lowest physical layer of the instruction set hierarchy for quantum computers~\cite{4rf7-9tfx}. 
Similarly, the choice of quantum error correcting codes that constitute the subsequent logical layers is closely intertwined with the understanding of the underlying quantum control methods. Having control protocols with rigorously bounded errors can provide the guarantee of success of such error correction techniques.
Consequently, the accuracy and optimality of quantum control methods are crucial factors in scaling quantum hardware to achieve utility. 

Quantum control makes use of time-dependent control pulses \cite{PhysRevLett.120.150401} to steer the system dynamics towards a desired target state. Therefore, designing these pulses with high precision is an important task. 
However, the current landscape of quantum control methods \cite{Koch2022}
is dominated by heuristic methods, such as gradient-based optimizers \cite{KHANEJA2005296, PhysRevLett.106.190501, PhysRevA.84.022326, Petruhanov_2023} and machine-learning-based approaches \cite{GIANNELLI2022128054, PhysRevX.12.011059, D2CP02495K, 11lg-bqbm}, where the optimality of the final solution is often not guaranteed, and the performance of the algorithm depends heavily on the choice of pulse-parameter initialization.  
These techniques often involve computationally intensive searches over the parameter space, rendering them opaque and offering limited understanding of the optimal solution. Therefore, it is desirable to design quantum control methods that can be analytically developed and understood.

Recently, there has been a growing interest in exploring analytic techniques to systematically design control pulses. One such continuous-time pulse design approach is the geometric space-curve formulation \cite{Barnes_2022,PhysRevA.99.052321,PhysRevA.108.012407,
amer2025implementingbenchmarkingdynamicallycorrected,piliouras2025automatedgeometricspacecurve}, which involves carefully designed pulses that cancel out various orders of undesirable interactions. In parallel, several other works \cite{PhysRevA.104.032605, Rainaldi2026} have developed analytic unitary constructions using the algebraic structure of Hamiltonians in qubit-oscillator systems. 
However, these analytical approaches typically seek closed-form solutions tailored to a specific problem under additional constraints.  For example, the standard formulation of the space-curve approach \cite{Barnes_2022,PhysRevA.99.052321,PhysRevA.108.012407,
amer2025implementingbenchmarkingdynamicallycorrected,piliouras2025automatedgeometricspacecurve} identifies time-dependent curves that satisfy particular constraints arising from differential equations. This inherently restricts the framework to single-qubit control, and extending it to multi-qubit settings becomes increasingly non-trivial. Furthermore, continuous-time control pulses with continuously varying amplitudes are often highly sensitive to instrumental noise, and fine-grained features of a smoothly varying waveform may be significantly degraded in low signal-to-noise regimes.

In this work, we pursue the direction of analytical quantum control, taking inspiration from quantum algorithms. Specifically, we adopt the quantum signal processing (QSP) algorithm~\cite{low2016methodology, low2019hamiltonian, PhysRevLett.118.010501} as the core control technique, and utilize the robustness of discretized pulse design and its ease of scalability to multi-qubit and qubit-oscillator systems. At a high level, QSP and its extensions \cite{PRXQuantum.5.020368, 10.1145/3313276.3316366, PRXQuantum.2.040203} are versatile tools for approximating a broad class of unitary transformations, and admit an analytical understanding while attaining high accuracy and bounded error. QSP achieves this by proposing a parameterized quantum circuit that outputs polynomial transformations of sub-blocks of an input unitary matrix. These polynomial transformations can be computed classically, tuned by the parameters in the circuit, and applied to a broad class of unitaries. Given this immense flexibility, QSP has found use in developing a wide variety of quantum algorithms, like quantum simulation, search, phase estimation, linear algebra, and beyond~\cite{PhysRevLett.118.010501, low2019hamiltonian, PRXQuantum.2.040203, Dong_2022_Ground, Wang_2025_EfficientGround}. Beyond algorithms, QSP has also been used to enhance metrology and device characterization~\cite{dong2025situquantum, Dong_2025_Optimal, marrero2026encoded}.

Here we focus on qubit-oscillator dynamics, which are extensively used in 
popular architectures such as superconducting \cite{RevModPhys.73.357, PhysRevA.69.062320, clarke2008superconducting, RevModPhys.93.025005} and trapped-ion quantum computers \cite{RevModPhys.73.565, doi:10.1126/science.1078446, Haroche2020}
to mediate qubit-qubit interactions during qubit gate implementations.  
We introduce a new framework, \emph{QSP-Control}, defined as a class of QSP protocols in which the signal operator is generated by native coupling of two distinct subsystems $A$ and $B$, and interleaved with programmable rotations on subsystem $A$. The resulting sequence implements polynomial transformations of the native couplings projected on to subsystem $B$, thereby enabling systematic control of subsystem $B$ dynamics toward a desired target state or operation. A particular instance of this framework is the qubit-oscillator systems, where native qubit-oscillator couplings act as the signal operators and programmable qubit rotations supply the control phases.

Using QSP-Control, we address the fundamental challenge of controlling qubit-oscillator dynamics in the presence of nonlinear Kerr interactions, which are a dominant source of error in superconducting transmon qubits without explicit hardware engineering~\cite{dgwk-w3jj}. In particular, after introducing necessary background in Sec.~\ref{sec:bg}, we present methods to effectively linearize the Kerr interactions in Sec.~\ref{sec:linearize}, thereby mitigating this source of noise. 
We show that achieving an error $\epsilon$ for a Fock level cutoff $N_{\rm {cutoff}}$ and resolution-dependent factor $w$ requires a circuit depth of 
$\mathcal{O}(\frac{1}{w} \log{(\frac{N_{\rm {cutoff}}}{ \epsilon})})$. 
Furthermore, in Sec.~\ref{sec:fock-manipulation}, we use QSP-Control to design constructions for selective Fock state manipulation of an oscillator. We achieve this by interleaving the Jaynes-Cummings \cite{Shore01071993} interaction (as is native to cavity-QED platforms) with qubit rotations to map this task to a QSP problem. As we highlight later in this paper, such selective control of Fock states facilitates the preparation of specialized bosonic states relevant to error-correcting codes \cite{PhysRevX.6.031006} and quantum memory \cite{t4cv-y398}. We conclude with a discussion in Sec.~\ref{sec:conclusion}.
Overall, this work offers a structured, analytic framework for quantum control, provides an alternative to heuristic and brute-force optimization methods, and demonstrates the far-reaching utility of the QSP framework.

\section{Background}\label{sec:bg}
As a brief introduction to the relevant background for this paper, we provide an overview on quantum oscillators and notation used in Sec.~\ref{sec:bg-notation}. We also include a concise introduction to quantum signal processing algorithms and their useful extensions in Sec.~\ref{subsec:bg-qsp}, with examples to familiarize the reader with the QSP framework.

\subsection{Notation for Quantum Harmonic Oscillators}\label{sec:bg-notation}

We will refer to the canonical position and momentum operators of harmonic oscillators as $\hat{x}$ and $\hat{p}$, respectively, which satisfy $[\hat{x}, \hat{p}] = i$, taking $\hbar = 1$. We consider these operators in the context of an oscillator with mass $m=1$, and frequency $\omega=1$.

We denote the bosonic annihilation and creation operators by $\hat{a}$ and $\hat{a}^{\dagger}$,  satisfying the relationship $[\hat{a}, \hat{a}^{\dagger}] = 1$. These operators are related to position and momentum as $\hat{a} = \frac{1}{\sqrt{2}}(\hat{x} + i\hat{p}), \ \hat{a}^{\dagger} = \frac{1}{\sqrt{2}}(\hat{x} - i\hat{p})$, or equivalently $\hat{x} = \frac{1}{\sqrt{2}}(\hat{a} + \hat{a}^{\dagger}), \ \hat{p} =-i\frac{1}{\sqrt{2}}( \hat{a} - \hat{a}^{\dagger})$. The photon number operator is expressed as $\hat{n} = \hat{a}^{\dagger}\hat{a}$. This acts on the oscillator Fock states, denoted by $\ket{r}$ for eigenvalues $r=0, 1, 2, \dots$, as
\begin{align}
    \hat{n}\ket{r} = r\ket{r}.
\end{align}
The photon number operator can also be expressed in terms of projectors onto the Fock states as
\begin{align}
    \hat{n} = \sum_{r=0}^{\infty} r \ket{r}\bra{r}. \label{eq:nhat-projectors}
\end{align}
Further, the action of the bosonic annihilation operator on a given Fock state $\ket{r}$ is
\begin{align}
    \hat{a}\ket{r} = \sqrt{r}\ket{r-1}, \quad  r\ge 1,
\end{align}
and that of the creation operator is given by
\begin{align}
     \hat{a}^{\dagger}\ket{r} = \sqrt{r+1}\ket{r+1}, \quad r\ge 0.
\end{align}

While these operators describe the infinite-dimensional Fock space of a bosonic mode, another system we will also study is a qubit, which is a two-level system. 
In this context, we refer to the qubit ground state and the excited state as $\ket{g}$ and $\ket{e}$, respectively, and the qubit Pauli operators as $\sigma_x, \sigma_y, \sigma_z$.

\subsection{Quantum Signal Processing}\label{subsec:bg-qsp}

Among the quantum algorithms developed in the past decades, quantum signal processing (QSP) \cite{low2016methodology, low2019hamiltonian, PhysRevLett.118.010501} stands out for its remarkabe versatility and its applicability to various problems in the quantum domain. 
The standard form of QSP interleaves two types of qubit rotation operators. These operators are the ``signal operator" that encodes a parameter $x \in [-1,1]$ of interest: 
\begin{align}
    W(x)  = \begin{pmatrix}
        x & i\sqrt{1-x^2} \\
        i\sqrt{1-x^2} & x
    \end{pmatrix} = e^{i\arccos(x) \sigma_x} , 
\end{align}
and a ``signal processing operator" that is parameterized by an angle $\phi \in[0,2\pi)$:
\begin{align}
    S(\phi) = e^{i \phi \sigma_z}.
\end{align}
Then, QSP considers a set of $d+1$ phase angles $\vec{\phi} = \{\phi_0, \phi_1, \ldots, \phi_d\} \in \mathbb{R}^{d+1}$ and proposes the following sequence of gates that interleaves $W(x)$ and $S(\phi_i)$:
\begin{align*}
    U_{\vec{\phi}}(x) &= S(\phi_0) \prod_{k=1}^{d} W(x) S(\phi_k) . 
\end{align*}
Remarkably, this sequence is proven to take the form:
\begin{align*}
    U_{\vec{\phi}}(x) &= \begin{bmatrix}
        P(x) & i Q(x)\sqrt{1-x^2} \\
        iQ^*(x)\sqrt{1-x^2} & P^*(x)
    \end{bmatrix}
\end{align*}
where, $P(x)$, and $Q(x)$ are polynomials parameterized by $\vec{\phi}$ that satisfy the following constraints
\vspace{-1mm}
\begin{itemize}
\setlength{\itemsep}{-.5mm}
    \item deg($P$) $\leq$ $d$, deg($Q$) $\leq$ $d-1$,
    \item parity of $P(x)$ is $d$ mod $2$, $Q(x)$ is $(d-1)$ mod $2$
    \item $|P(x)|^2 + (1-x^2)|Q(x)|^2 = 1$, $\forall$ $x \in [-1,1]$.
\end{itemize}
This result tells us that the sequence $U_{\vec{\phi}}$ realizes a polynomial transformation of the signal $x$. Moreover, these polynomials can be classically simulated and analytically understood. In fact, for any desired polynomial that obeys the above conditions, a corresponding set of phases can be found classically in $\text{poly}(d)$ time~\cite{Haah2019product, chao2020findingangles, Dong_2021_Efficient, yamamoto2024robustangle, Ying2022stablefactorization}. The cost of a QSP implementation, which is the number of queries to the signal operator $W$, is found to be $d$, where $d$ is the degree of the target polynomial~\cite{PhysRevLett.118.010501}.

While QSP is a general and flexible framework, the polynomials it can produce are limited by the aforementioned constraints. A particularly restrictive constraint is that $P$ and $Q$ must have definite parity (either be even or odd). To alleviate this, Ref.~\cite{PRXQuantum.5.020368} proposed Generalized Quantum Signal Processing (GQSP). In GQSP, the signal processing operator is replaced by an arbitrary $SU(2)$ rotation gate $R(\theta, \phi, \lambda)$ 

\begin{align}
     R(\theta, \phi, \lambda) = \label{eq:gqsp-rot}
\begin{bmatrix}
    e^{i(\lambda+\phi)}\cos{(\theta)} & e^{i\phi}\sin{(\theta)}
    \\
    e^{i\lambda}\sin{(\theta)} & -\cos{(\theta)}
\end{bmatrix}.
\end{align}
This generalization lifts the parity constraints imposed by QSP and can realize Laurent polynomials with complex coefficients, where a Laurent polynomial is a polynomial of the form $\sum_{k\in \mathbb{Z}} a_k x^k$ defined over a complex field.  

Moreover, the polynomial transformations provided by GQSP can be generalized to an eigenvalue transformation, that acts on the eigenvalues of a Hermitian operator. Specifically, take $A\in \mathbb{C}^{2^n\times 2^n}$ to be a Hermitian matrix, suppose we are given access to a controlled unitary transformation $cU(A) \in \mathbb{C}^{2^{n+1}\otimes 2^{n+1}}$ of the form
\begin{equation}
    cU(A):= \begin{bmatrix}
        \openone & 0 \\0& e^{iA} 
    \end{bmatrix} , 
\end{equation}
which block-encodes $e^{iA}$. Given similar access to a block-encoding of $e^{-{iA}}$,
we can construct a block-encoding $U'_Z(A)$ of the form
\begin{equation}
    U'_Z(A) = \begin{bmatrix}
        e^{-i A} &0\\
        0 & e^{iA}
    \end{bmatrix} , 
\end{equation}
which upon conjuagation by a Hadamard gate to the control qubit becomes
\begin{equation} \label{eq:general-Z-to-X-basis-conv}
    U'_X(A) = \begin{bmatrix}
        \cos(A) & -i\sin(A) \\
        -i\sin(A) & \cos(A)
    \end{bmatrix}. 
\end{equation}
This resembles an $X$-rotation and serves as an analog of the signal operator $U(x)$,  where in this context $x = \cos(A)$.  Similarly, the block matrix
\begin{equation}
    \begin{bmatrix}
        e^{-i\theta} \openone & 0 \\ 0 & e^{i\theta} \openone
    \end{bmatrix}
\end{equation}
resembles a $Z$-rotation and serves as an analog of the signal processing operator. It can be implemented using a single-qubit $R_z(\theta)$ gate on the control qubit. By interspersing these operators as in GQSP, this generates polynomials in $A$ (acting on its eigenvalues) and allows us to perform GQSP on a block encoding of a unitary transformation without substantial modification.

It should be noted that QSP and GQSP are the simplest forms of this framework, restricted to the design of single variable polynomials. However, the scope of QSP extends well beyond this setting. Recent works have introduced multivariate generalizations that encompass commuting \cite{Rossi2022multivariable} and non-commuting variables \cite{singh2025nonabelianquantumsignalprocessing,11129874}, as well as extensions to groups beyond SU(2) \cite{Lu2026quantumsignal,laneve2024quantumsignalprocessingsun, rossi2023quantumsignal}. Moreover, QSP admits a range of further generalizations to modular constructions~\cite{Rossi2025modularquantum,gomes2024multivariableqspbosonicquantum}, infinite degree sequences \cite{Dong2024infinitequantum}, parallel implementations \cite{Martyn2025parallelquantum, Zhang2024parallelquantum, goldsteingelb2025compasdistributedmultipartyswap}, and realizations as a randomized algorithm~\cite{Martyn_2025, Wang_2026_Random}. Together, these developments expand the expressive power of QSP, while addressing practical challenges such as noise, circuit depth, and hardware constraints~\cite{10.1063/5.0312254}. Although we restrict our attention here to standard single variable QSP, these advances position the broader QSP framework as a versatile toolbox for exploring quantum control.

\section{Linearizing Kerr Interactions in Circuit-QED with QSP-Control}
\label{sec:linearize}
In this section, we first review the origin of the Kerr nonlinearity in superconducting qubits coupled to microwave cavities (oscillators) in Sec.~\ref{subsec:Linearize-bg} and then introduce the main problem we solve in Sec.~\ref{subsec:Linearize-main-problem}. We explain our approach to solving the problem in Sec.~\ref{subsec:Linearize-gqsp-formulation} and discuss the error bounds in Sec.~\ref{subsec:Linearize-error-scaling}.

\subsection{Background on Superconducting Circuit-QED}\label{subsec:Linearize-bg}

In superconducting quantum computing platforms, the most widely used form of qubit, known as the transmon qubit, is engineered from Josephson junction circuits \cite{PhysRevA.76.042319, Houck2009, Babu2021, Willsch2024}. A practical Josephson junction introduces a nonlinear potential energy that depends on the cosine of the superconducting phase difference $\hat{\phi}$ across the junction, which can be effectively expressed as, 
\begin{align}
    \hat{H}_J  \propto E_J \cos{(\hat{\phi})}.
\end{align}
Here, $E_J$ is the Josephson energy of the fabricated Josephson junction device. Taylor expanding this cosine potential yields a leading quadratic term representing a standard harmonic oscillator, and higher-order quartic term, which gives rise to the system's anharmonicity.

Hybrid qubit-oscillator systems can be designed by coupling a transmon qubit to microwave cavities, which behave as quantum harmonic oscillators \cite{PhysRevLett.108.240502}. Universal control on qubit-oscillator systems \cite{PhysRevLett.76.1055, PhysRevA.92.040303, Eickbusch2022} has been shown to be achieved by establishing a dispersive coupling interaction between the qubit and the cavity, with a desired Hamiltonian of the form, 
\begin{align}
    \hat{H} = \chi \hat{n} \sigma_z , \label{eq:desirable-hamil-lin}
\end{align}
where $\chi$ is the dispersive coupling factor. 
However, the anharmonicity of a transmon qubit impacts the dynamics of the system when coupled to a microwave cavity, leading to effects such as qubit-photon number-dependent shifts, also known as the cross-Kerr interaction, and cavity photon-photon interactions, known as the self-Kerr \cite{PhysRevA.86.013814, RevModPhys.93.025005, PhysRevA.84.012329, PhysRevA.90.023827, Kounalakis2018}. With these effects taken into account, the overall physical Hamiltonian 
becomes, 
\begin{align}
    \hat{H}_{\rm eff} 
    = \label{eq:kerr-hamitonian}
    \chi \sigma_z \hat{n} + k' \sigma_z \hat{n}^2 + k \hat{n}^2,
\end{align}
where $k$ and $k'$ are the self- and cross-Kerr coefficients, respectively. 
The combined effects of cross- and self-Kerr interactions in qubit-oscillator systems give rise to state-dependent nonlinear phase accumulation, which causes dephasing of the qubits, gate errors, and reduced fidelity of qubit-oscillator operations.

The Selective Number-dependent Arbitrary Phase (SNAP) gate was introduced in Ref.~\cite{PhysRevLett.115.137002} to mitigate such unwanted phase evolution caused by Kerr interactions \cite{2020efficientcavitycontrolsnap, smcc-t465, PhysRevLett.101.240401}. On a qubit-oscillator system, the SNAP gate is expressed as,
\begin{align}
    \text{SNAP}(\varphi) = e^{-i\sum_n \sigma_z \varphi_n \ket{n}\bra{n}}, \quad \varphi_n \in [0, 2\pi]
\end{align}
$\forall~n \in \mathbb{Z}^+$, where $\varphi_n$ are individually programmable phases associated with each Fock level $\ket{n}$, enabling selective control of number-dependent phase evolution.
This operation has since been shown to provide efficient control of circuit-QED when used in conjunction with oscillator displacement gates.
However, in practice, the SNAP gate is often limited by the spectral selectivity required to resolve number-dependent transitions associated with different Fock levels. As the photon-number range increases, maintaining this selectivity generally requires longer gate durations, thereby introducing a trade-off between control fidelity and decoherence. Furthermore, constructing a SNAP gate that accurately implements arbitrary number-dependent phases becomes computationally challenging at large photon-number cutoff, because realizing arbitrary unitaries requires $\mathcal{O}(n_{\text{max}}^2)$ operations, where $n_{\text{max}}$ denotes the maximum photon number~\cite{PhysRevLett.115.137002}.
For these reasons, our aim in this section is to develop quantum signal processing into a scalable methodology for controlling qubit-oscillator systems and mitigating the effects of Kerr interactions with reduced implementation costs.

\subsection{Linearization Problem}\label{subsec:Linearize-main-problem}

We consider the qubit-oscillator Hamiltonian with second-order Kerr nonlinearities given in Eq.~\eqref{eq:kerr-hamitonian}. Among the nonlinear contributions, the cross-Kerr term couples to the qubit state, whereas the self-Kerr term is independent of it. As a result, only the cross-Kerr contribution can be directly modulated through qubit control. We therefore isolate the qubit-state-dependent part of Eq.~\eqref{eq:kerr-hamitonian} and consider a physical Hamiltonian affected only by the cross-Kerr term:
\begin{align}
     \hat{H}' 
    = \label{eq:only-crosskerr-hamitonian}
    \chi \sigma_z \hat{n} + k' \sigma_z \hat{n}^2 .
\end{align}
We express the unitary operator generated by time evolution under $\hat{H}'$ for a time duration $\tau$ as ${\rm CR'}(\tau)$:
\begin{align}
    {\rm CR'}(\tau) &= \nonumber
    e^{-i \hat{H}' \tau} 
    = 
   e^{-i \tau(\chi  \hat{n} + k' \hat{n}^2)\sigma_z} 
    \\
    &= 
    \begin{bmatrix}
        e^{-i \tau (\chi \hat{n} + k' \hat{n}^2 )} & 0 \\
        0 & e^{i \tau (\chi \hat{n} + k' \hat{n}^2)}
    \end{bmatrix}.
    \label{eq:CR'}
\end{align} 
In practice, we wish to mitigate the effects of the Kerr interactions and instead evolve under the desired Hamiltonian $\hat{H}$ shown in Eq.~\eqref{eq:desirable-hamil-lin}. 
Let us denote the unitary corresponding to evolution under this Hamiltonian for a  time duration $\tau$ by ${\rm CR}(\tau)$, which takes the form, 
\begin{align}
    {\rm CR}(\tau) 
    = \label{eq:CR}
    e^{-i\tau \chi \sigma_z \hat{n}}.
\end{align}
By expanding $\hat{n}$ in the Fock basis per Eq.~\eqref{eq:nhat-projectors}, ${\rm CR}(\tau)$ can be expressed as a sum of operations acting on subspaces spanned by the joint qubit-oscillator states $\ket{g,r}$ and $\ket{e, r}$, 
\begin{align}\label{eq:CR_tau}
    {\rm CR}(\tau) =  \sum_{r=0}^{\infty}
    \begin{bmatrix}
       e^{-i\tau \chi r}  & 0
        \\
        0 &  e^{i\tau \chi r }
    \end{bmatrix} \otimes \ket{r}\bra{r} , 
\end{align}
where $\ket{r}$ is the Fock number for Fock level $r$.
Similarly, ${\rm CR'}(\tau)$ can be expressed as, 
\begin{align}
    &{\rm CR'}(\tau) = \sum_{r=0}^{\infty}
   \begin{bmatrix}
       e^{-i \tau (\chi r + k' r^2 )} & 0 \\
        0 & e^{i \tau (\chi r + k' r^2)}
    \end{bmatrix}\otimes\ket{r}\bra{r} .
\end{align}
As we mentioned earlier, our desired Hamiltonian is free of the cross-Kerr term $k'\hat{n}^2\sigma_z$. 
Thus, we pose as our main problem the following task of linearizing the Hamiltonian: 
\\
\\
\textbf{Linearization Problem}: 
\emph{
Consider a qubit-oscillator system with Kerr nonlinearities, where the oscillator occupation number is truncated at $N_{\rm cutoff} \in \mathbb{Z}^+$, and the native Hamiltonian $\hat{H}'$ is given in Eq.~\eqref{eq:only-crosskerr-hamitonian}. Assuming access to arbitrary SU(2) qubit rotations as defined in Eq.~\eqref{eq:gqsp-rot}, our goal is to emulate the evolution under a target Hamiltonian $\hat{H}$ given in Eq.~\eqref{eq:desirable-hamil-lin}, over a total time $\tau$ within an error $\epsilon$ by using the qubit rotations and the native Hamiltonian $\hat{H}'$. 
}

In the following subsection, we use the QSP-control framework to eliminate the impact of Kerr nonlinearity, effectively realizing a linearized form of $\hat{H}'$ that closely approximates the desired Hamiltonian $\hat{H}$ of Eq.~\eqref{eq:desirable-hamil-lin}. It is to be noted that for practical systems, this linearization can only be obtained for a finite range of Fock levels due to physical limitations. Thus, we have imposed an upper limit of $N_{\rm {cutoff}}$ as the highest Fock level up to which the Hamiltonian is linearized.

\subsection{Linearizing Kerr Interactions with QSP}
\label{subsec:Linearize-gqsp-formulation}

To motivate our approach and connect it to QSP, we first rewrite the target operation ${\rm CR}(\tau)$ in a form amenable to polynomial approximation. From Eq.~\eqref{eq:CR_tau}, ${\rm CR}(\tau)$ corresponds to a qubit $Z$-rotation with a Fock-number-dependent angle. Conjugating by a Hadamard gate $H$ transforms this into an $X$-rotation, yielding
\begin{align}
  &H {\rm CR}(\tau)H 
  = \nonumber
  H\begin{bmatrix}
        e^{-i \tau \chi \hat{n}} & 0 \\
        0 & e^{i \tau \chi \hat{n}}
    \end{bmatrix} H =  e^{-i\tau \chi \sigma_x \hat{n}}
    \\
    &= \label{eq:fock-basis} 
    \sum_{r=0}^{\infty}
    \begin{bmatrix}
        \cos{(\tau \chi r)}  & -i\sin{(\tau \chi r)} 
        \\
        -i\sin{(\tau \chi r)}  &  \cos{(\tau \chi r)} 
    \end{bmatrix}\otimes\ket{r}\bra{r}.
\end{align}
In this representation, our target unitary decomposes into a direct sum of $X$ rotations, where each Fock level $r$ experiences a rotation through an angle $\tau \chi r$. 
We can approximate the unitary shown in Eq.~\eqref{eq:fock-basis} by using GQSP to approximate the function $\cos(\tau \chi r)$ over the range $r \in [0, N_{\rm {cutoff}}]$. We will do so using GQSP, which affords greater flexibility in the functions we can approximate. 
In addition, by approximating $\cos(\tau \chi r)$, the $\sin(\tau \chi r)$ terms are automatically reproduced up to a gauge phase due to the unitarity of the GQSP construction.

We now describe the core construction of the GQSP used to address the linearization problem. Our approach is to interleave short-time evolution of duration $t$, under the qubit-oscillator entangling gate ${\rm CR'}(t)$ with single-qubit rotations, thereby mapping the resulting dynamics onto a GQSP sequence. Specifically, by composing $d$ such layers with a total circuit runtime of $t d$, we obtain the total unitary 
\begin{align}
    &U_d
    =
    \left[
    \prod_{j=1}^d R(\theta_j, \phi_j, 0) {\rm CR'}(t)
    \right]  R(\theta_0, \phi_0, \lambda_0)
    \\
    &= \label{eq:kerr-lin-gqsp-poly-blk}
    \begin{bmatrix}
    P(e^{-i t (\chi \hat{n} + k' \hat{n}^2 )} ) &  -Q^{\dagger}(e^{-i t (\chi \hat{n} + k' \hat{n}^2 )} )
        \\
        Q(e^{-i t (\chi \hat{n} + k' \hat{n}^2 )} ) 
        & P^{\dagger}(e^{-i t (\chi \hat{n} + k' \hat{n}^2 )} )
    \end{bmatrix}
\end{align}
where $P$ and $Q$ are the GQSP polynomials. 
We will use this unitary to approximate ${\rm CR}(\tau)$ as in Eq. \eqref{eq:fock-basis}. 
Ref.~\cite{PhysRevA.110.012612} proves that $P$ and $Q$ are degree $d$ Laurent polynomials in $e^{-i t (\chi \hat{n} + k' \hat{n}^2 )}$, each with fixed parity $d \mod 2$. Explicitly, they are given by  
\begin{align}
    P (e^{-i t (\chi \hat{n} + k' \hat{n}^2 )} )
    &= 
    \sum_{m=-d}^{d} p_m e^{-i m t (\chi \hat{n} + k' \hat{n}^2 )} ,
    \\
    Q (e^{-i t (\chi \hat{n} + k' \hat{n}^2 )} )
    &= 
    \sum_{m=-d}^{d} q_m e^{-i m t (\chi \hat{n} + k' \hat{n}^2 )} 
\end{align}
for coefficients $p_m, q_m \in \mathbb{C}$ with $m= \{-d, -d+2, \cdots, d-2, d\}$ \cite{PhysRevA.110.012612}.
By joint optimization over the polynomials $P$ and $Q$, it is possible to realize the desired phase value corresponding to each Fock level by appropriate choice of GQSP phase angles.

Our goal here is to match Eq.~\eqref{eq:fock-basis} by using the QSP sequence in Eq.~\eqref{eq:kerr-lin-gqsp-poly-blk} which are equivalent up to Hadamard conjugation.
Note that $\tau$ here does not have to be equal to $t d$, in general.
For each Fock level $r$, the desired polynomial approximation can be alternatively expressed as,
\begin{align}
    P(e^{-i t (\chi r + k' r^2 )}) &\approx  \cos{(\tau\chi r)}
    \\
     Q(e^{-i t (\chi r + k' r^2 )}) &\approx -i \sin{(\tau\chi r)} . 
\end{align} 
This expression highlights that the overall construction realizes a Fock-level-dependent SU(2) rotation that approximates the desired evolution.

\begin{figure}[ht]
    \centering
    \includegraphics[width=1\linewidth]{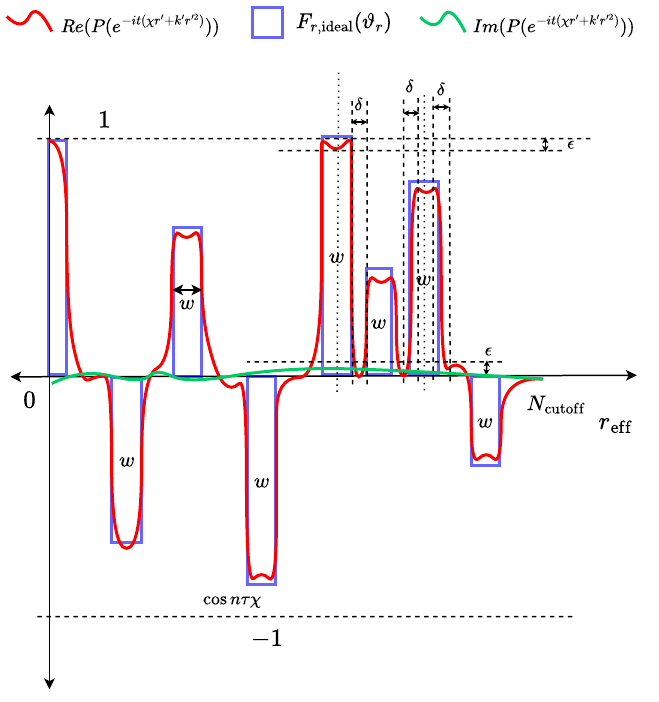}
    \vspace*{-1\baselineskip}
    \caption{Illustration of the polynomial approximation of phase values for ${\rm CR}(\tau)$. The schematic shows how the distinct cosine values corresponding to the various Fock levels are approximated. The height of the blue boxes are the ideal $\cos(\tau \chi r)$ values that the protocol intends to approximate.}
    \label{fig:ideal-linearization}
\end{figure}

Given that the desired linearization must hold for all Fock levels up to $N_{\rm {cutoff}}$, the problem reduces to designing a QSP polynomial $P$ such that it closely approximates $\cos(\tau \chi r)$ for each $r \in \{0, 1, \cdots N_{\rm {cutoff}}\}$, within a prescribed error tolerance.
It should be noted that the underlying variable of the polynomial is $e^{-i t (\chi r + k' r^2 )} = e^{-i t \chi r_\text{eff}}$ with the effective Fock number as 
\begin{align}
    r_{\rm eff} = r \left(1 + \frac{k'}{\chi}r \right) ,
\end{align}
which need not be an integer.
The equivalent fractional Fock level $r_{\rm eff}$ is non-uniformly distributed in the range $[0, N_{\rm cutoff}]$ and may become arbitrarily close for some integer Fock levels. Further, for all Fock levels $n$, for which,
\begin{align}
    n >  \frac{2\pi}{\chi \tau},
\end{align}
a wrap-around of the Fock number occurs.
We seek to approximate $P(e^{-i t (\chi r + k' r^2 )})$ as a function of $r_{\rm eff}$
within a small interval of width $w$ around the value $r$ for all Fock level of interest, as can be seen in Fig.~\ref{fig:ideal-linearization}.
This requires 
\begin{align}
&\Re(P(e^{-i  t \chi r_{\rm eff}})) \nonumber
    \approx \begin{cases}
      \cos(\tau \chi r),  ~~ r \in [0, N_{\rm cutoff} ]
      \\
      0,\quad  \text{otherwise}
    \end{cases}
   \\
   &
    \Im(P(e^{-i  t \chi r_{\rm eff}})) \approx 0,\quad \forall r.
\end{align} 
The choice of $w$ is determined by the minimum separation between distinct values of $r_{\text{eff}} $ after wrap-around, over the truncated Fock space, ensuring that these intervals remain non-overlapping and thus distinguishable. 
That is, we will set
\begin{align}
    w < \min_{r \neq r'} \left| r_{\rm eff} - r'_{\rm eff} \right|
\end{align}
as the minimum distance between distinct $r_{\rm eff}$
values over the truncated Fock space for  $r = 0, 1, 2, \cdots, N_{\rm cutoff}$, ensuring that the corresponding intervals do not overlap. 
Therefore, the GQSP construction aims to realize a polynomial transformation such that 
$\Re(P(e^{-i  t \chi r_{\rm eff}}))$ approximates $\cos(\tau \chi r)$ for each Fock level.
For a fixed Fock level $r \in \{0, 1, \cdots, N_{\rm cutoff}\}$, we define an \emph{ideal target function} for the polynomial approximation as,
\begin{align}
    F_{r,\text{ideal}}(r_{\rm eff}) = 
    \begin{cases}
         \cos(\tau \chi r), 
          ~~
         r_{\rm eff} \in [r-\frac{w}{2}, r+\frac{w}{2}],
        \\
        0, 
        \quad \quad 
        \text{otherwise}.
    \end{cases}
\end{align}
The above expression can be equivalently written using a top-hat function as
\begin{align}
    F_{r, \text{ideal}}(r_{\rm eff}) = \cos(\tau \chi r)\, F_{\text{Hat}}\!\left(r_{\rm eff} - r\right),
\end{align}
where
\begin{align}
    F_{\text{Hat}}(x) = 
    \begin{cases}
        1, & x \in [-w/2,\, w/2], ~x\in \mathbb{R}\\
        0, & \text{otherwise}.
    \end{cases}
\end{align}
To simultaneously accommodate all Fock levels up to $N_{\rm {cutoff}}$, 
we piece together all the approximations from the individual Fock levels to give us the overall target function as,
\begin{align}
    F_{\text{ideal}}(r_{\rm eff}) = \sum_{r=0}^{N_{\rm {cutoff}}} F_{r, \text{ideal}}(r_{\rm eff}),
\end{align}
which corresponds to a superposition of top-hat functions centered at the Fock number $r$, each with height $\cos(\tau \chi r)$ and width $w$. The goal is then to construct a GQSP polynomial that approximates this target function over the relevant domain.
As depicted in Fig.~\ref{fig:ideal-linearization}, the width  $w$ of the neighborhood around which we map the value of $r_{\rm eff}$, is shown to be the width of the \emph{ideal target function} box (blue colour) corresponding to each Fock level. We choose $w$ such that these ideal blue boxes do not overlap. The red and the green curves indicate the real and imaginary parts of $P(e^{-i  t \chi r_{\rm eff}})$ respectively.

Since a discontinuous step or top-hat function cannot be realized exactly by a finite-degree polynomial, we instead approximate it by allowing a small transition region around each discontinuity where the polynomial approximation fails. We denote the width of this transition region by $\delta$ as shown in Fig.~\ref{fig:ideal-linearization}, and take $\delta \leq w$ to ensure a correct approximation of the target function at the Fock levels up to $N_{\text{cutoff}}$. Thus, $\delta$ defines a buffer region between adjacent intervals, enabling smooth polynomial interpolation and reducing approximation ripples near the boundaries. The precise choice of $w$ and $\delta$ governs the resolution and robustness of the approximation; we discuss these choices and their impact on the polynomial complexity in the next section. 

\subsection{Polynomial Construction and Complexity}\label{subsec:Linearize-error-scaling}

To analyze the trade-off between the required polynomial degree and the approximation error, we begin by considering a single target function $F_{r,\mathrm{ideal}}(r_{\rm eff})$. Each such function is a scaled and shifted top-hat function centered at Fock level $r$. Since a discontinuous top-hat function cannot be represented exactly by a finite-degree polynomial, we instead allow a narrow transition region of width $\delta$ around each discontinuity, within which the polynomial approximation deviates from the ideal function. Outside this transition region, the approximation is required to remain within error $\varepsilon$. Using the constructions of Refs.~\cite{low2017hamiltoniansimulationuniformspectral,10.1145/3313276.3316366}, such a top-hat function can be approximated by a polynomial of degree
\begin{align}
    \mathcal{O}\!\left(\frac{1}{\delta}\log\frac{1}{\varepsilon}\right).
\end{align}
For our case, the complete ideal target function is obtained by summing $N_{\text{cutoff}}$ such top-hat functions, one for each Fock level up to the cutoff. Therefore, to approximate the full target within overall error $\epsilon$, it suffices to approximate each individual top-hat function to error $\varepsilon = \epsilon / N_{\text{cutoff}}$. This corresponds to a requisite polynomial degree
\begin{align}
    d = \mathcal{O}\!\left(\frac{1}{\delta}\log\!\left(\frac{N_{\text{cutoff}}}{\epsilon}\right)\right).
\end{align}
Finally, relating the transition width $\delta$ to the minimum resolvable separation $w$ between neighboring target cosine values, we require $\delta \leq w$ and thus obtain
\begin{align}
     d = \mathcal{O}\!\left(\frac{1}{w}\log\!\left(\frac{N_{\text{cutoff}}}{\epsilon}\right)\right),
     \label{eq:d-linearization-complexity}
\end{align}
which is the required polynomial degree for approximating the full target function.

It is important to note that the transition width $\delta$ cannot be chosen freely, but rather has dependency on $N_{\rm {cutoff}}$ and $w$ because both are constrained by the finite range of the target function. 
In particular, because the Fock levels lie in the interval $[0, N_{\rm cutoff}]$,
then a sufficient packing condition for $N_{\rm {cutoff}}$ rectangular bands of width $w$, separated by transition regions of width $\delta$, is
\begin{align}
    N_{\rm {cutoff}} w + (N_{\rm {cutoff}} - 1)\delta \le  N_{\rm {cutoff}}.
\end{align}
Because the nonlinear terms vanish at $r=0$, we impose the packing condition only on the Fock levels $r=1,\ldots,N_{\rm cutoff}$.
More generally, the relevant packing condition is governed by the minimum separation between the distinct values of the $r_{\rm eff} $.

\begin{figure}[ht]
    \centering
    \includegraphics[width=1\linewidth]{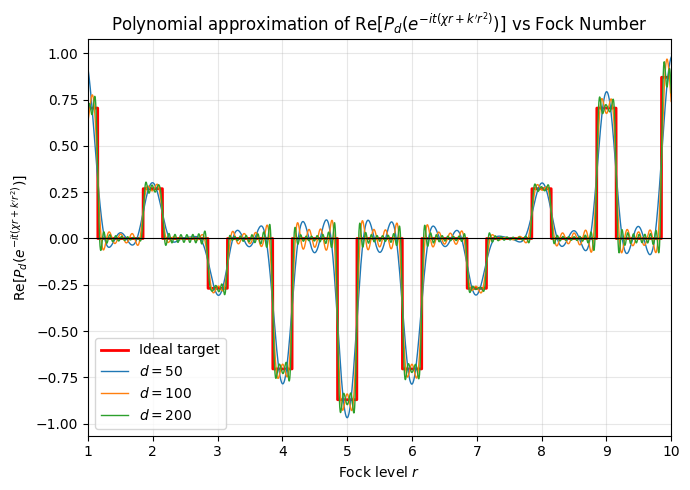}
    \vspace*{-1\baselineskip}
    \caption{Polynomial approximations to the phase values of ${\rm CR}(\tau)$ for polynomial degrees $d=50$, $100$, and $200$, evaluated over Fock levels $r=1,\ldots,10$. The results are obtained using $\chi/2\pi = 3~\mathrm{MHz}$, $k/2\pi = 5.6~\mathrm{kHz}$, $\tau = 2.7~\mu\mathrm{s}$, and time $t=18~\mathrm{ns}$. We choose a window width of $w=0.3$ and a transition width of $\delta=w/2$. The total circuit runtime is $t d$, where $d$ is the polynomial degree.}
    \label{fig:linearization-sim}
\end{figure}

Fig.~\ref{fig:linearization-sim} shows simulation results for approximating the cosine values of the linear phases, $\cos(\chi \tau r)$, for Fock levels $1$ through $10$. It can be observed that increasing the polynomial degree reduces the ripples in the polynomial approximation near the zeros and target values of the ideal function. In practical settings where qubit coherence times are limited, employing a long QSP sequence with time-step size $t$, such that each step are much smaller as compared with the qubit $T_1$ and $T_2$ times~\cite{10.1063-1.5089550}, may help reduce the effect of qubit under- or over-rotation errors due to decoherence on the overall polynomial approximation.

\section{Selective Manipulation of Fock States with QSP-Control}
\label{sec:fock-manipulation}

Several previous works \cite{PhysRevA.63.012306,PhysRevLett.95.010504, PhysRevA.87.022341, PhysRevA.104.032605} have theoretically established that the Jaynes-Cummings interaction, along with qubit rotation gates, are capable of realizing an arbitrary unitary on a truncated oscillator subspace. Moreover, Ref.~\cite{um2015realization} realized experimentally similar approximate phonon arithmetic deterministically via adiabatic passage. However, the optimality of such approaches, along with boundedness of the approximation error, have not been established methodically. In this section, we use QSP-Control to selectively add or remove photons from chosen Fock-state components.

\subsection{Motivation and Definition of the Problem}
As an example, consider a superposition of Fock states,
\begin{align}
    \ket{\psi}_{\mathrm{osc}} = \label{eq:general-sup-osc-state}
    c_1\ket{r_1} + c_2\ket{r_2}, \quad c_i \in \mathbb{C},~~ r_i \in \mathbb{Z}^+,
\end{align}
with normalization $|c_1|^2 + |c_2|^2 = 1$. Applying the bosonic creation operator $\hat{a}^\dagger$ to $\ket{\psi}_{\mathrm{osc}}$ uniformly increases the photon number of the constituent Fock states by 1,
\begin{align}
    \hat{a}^\dagger \ket{\psi}_{\mathrm{osc}} 
    &= 
    c_1 \sqrt{r_1+1}\ket{r_1 + 1}
    + c_2 \sqrt{r_2+1}\ket{r_2 + 1}.
\end{align}
This illustrates that the bosonic creation and annihilation operators act globally on the state and cannot be restricted to act on individual Fock levels. Here we fill this gap.

As another example, consider the Jaynes-Cummings interaction \cite{Shore01071993}, which is native to cavity-QED platforms. This is described by the expression, 
\begin{align}
    {\rm JC}(\theta, \varphi) &= \label{eq:JC-hamiltonian}
    \exp[-i\theta(e^{i\varphi}\sigma_{-}a^{\dagger} + e^{-i\varphi}\sigma_{+}a)] , 
\end{align}
where $\sigma_{-} = \ket{g}\bra{e}$ and $\sigma_{+} = \ket{e}\bra{g}$ are qubit transition operators acting on the qubit ground and excited states $\ket{g}, \ket{e}$, respectively.
In this basis, $\sigma_z = \ket{g}\bra{g} - \ket{e}\bra{e}$ is the Pauli-$Z$, and
$\sigma_x = \ket{g}\bra{e} + \ket{e}\bra{g}$ is the Pauli-$X$. 

Evidently, the Jaynes-Cummings coupling transfers excitations between the qubit and the oscillator while conserving the total excitation number. Depending on the value of $\theta$ and the oscillator Fock level involved, the excitation transfer for a given transition can be complete or partial. The direction of excitation transfer depends on the initial state of the qubit. When the qubit is in the ground state $\ket{g}$, it can absorb an excitation from the oscillator, thereby decrementing the photon number of each populated Fock component in the superposition. In contrast, when the qubit is in the excited state $\ket{e}$, it can emit an excitation into the oscillator, thereby incrementing the photon number of each populated Fock component.
However, achieving selective control over a particular Fock state is nontrivial, because the Jaynes-Cummings interaction acts on the entire oscillator state through the action of $\hat{a}$ and $\hat{a}^\dagger$ operators, thereby simultaneously coupling all populated Fock states in the superposition.

Consequently, there is a need for unitary operations that enable selective photon number manipulation, for instance, mapping $\ket{r_2} \rightarrow \ket{r_2+1}$ while leaving all other Fock components ($\ket{r_1}$ in the above example shown in Eq.~\eqref{eq:general-sup-osc-state}) unchanged.
Moreover, if an operator maps a Fock state $\ket{r_2}$ to $\ket{r_2\pm1}$, then cascading multiple such operators can realize mappings of the form $\ket{r_2}\mapsto\ket{r_2\pm m}$, where $m\in\mathbb{Z}^{+}$.
Such selective control is valuable in applications including the preparation of NOON states \cite{wfj8-tgjz}, the construction of bosonic error-correcting binomial codes \cite{PhysRevX.6.031006}, and the encoding of qudits in bosonic oscillators \cite{PhysRevA.87.022341}, where precise manipulation of individual Fock levels can significantly enhance control and functionality. Therefore, our specific problem is to selectively manipulate Fock states as follows: 
\\
\\
\\
\textbf{Selective Fock-State Manipulation Problem}: \emph{Consider an oscillator state as a superposition of Fock states $|\psi\rangle_{\mathrm{osc}} = \sum_i c_i |r_i\rangle$. Given a qubit-oscillator system with Jaynes--Cummings coupling and access to qubit rotation gates, our goal is to realize Fock-number-selective operations that add or subtract a photon from a chosen Fock state $|r\rangle$ in $|\psi\rangle_{\mathrm{osc}}$, while leaving all other populated Fock components unchanged. Here, we assume that the target level $\ket{r\pm 1}$ is not initially populated. This avoids collisions in which the selectively shifted component overlaps with an already occupied Fock state.}

\subsection{Intuition on Mapping the Jaynes-Cummings Gate to QSP}\label{subsec:qsp-ctrl}

The Jaynes-Cummings interaction shown in Eq.~\eqref{eq:JC-hamiltonian} can be decomposed into a direct sum of operators acting on 2$\times$2 subspaces, each spanned by $\{ \ket{e, n}, \ket{g, n+1}\}$ in the joint qubit-oscillator system state. Explicitly, in such a subspace, it can be expressed as
\begin{align}
    &{\rm JC}(\theta, \varphi)|_{(n, n+1)}  \nonumber
    \\
    &= \label{eq:JC-sub-block}
    \begin{bmatrix}
    \cos(\theta \sqrt{n+1}) & -i \sin(\theta \sqrt{n+1}) e^{i\varphi} \\
    - i \sin(\theta \sqrt{n+1}) e^{-i\varphi} & \cos(\theta \sqrt{n+1})
    \end{bmatrix}
    \\
    &= \label{eq:JC-as-QSP-block}
    e^{i\frac{\varphi}{2}\sigma_z}
    \begin{bmatrix}
    x_{n, n+1} & -i \sqrt{1-x_{n, n+1}^2}  \\
    - i \sqrt{1-x_{n, n+1}^2}   & x_{n, n+1}
    \end{bmatrix}
    e^{-i\frac{\varphi}{2}\sigma_z}, 
\end{align}
where $x_{n, n+1} = \cos(\theta \sqrt{n+1})$.
Thus, the Jaynes-Cummings gate couples adjacent Fock levels of the oscillator with qubit states, and hence admits this decomposition.
It can be observed in Eq.~\eqref{eq:JC-as-QSP-block} that in each of these blocks, the $Z$-rotation angle $\varphi$ is the same, but $X$-rotation angle $\theta\sqrt{n+1}$ is different, depending on the Fock number $n$. By this logic, we see that a Jaynes-Cummings interaction resembles an interleaved term in a QSP sequence, where the analogues of the signal is the value $\cos(\theta\sqrt{n+1})$ and the phases are $\varphi$.

Given this similarity to QSP, we propose a formulation by composing several Jaynes-Cummings interactions together and interpret the results from a polynomial transformation perspective. In practice however, it could be a complex task to manipulate the angle $\varphi$ in a Jaynes-Cummings interaction because this parameter is hardware-constrained.
We therefore include an additional qubit rotation $R_z(\phi)$ that absorbs the effect of the $\varphi$ rotation and affords us better control over the QSP sequence. Therefore, by introducing a set of angles $\vec{\phi} = \{\phi_0,\phi_1, \cdots,\phi_d\}$, we arrive at the following sequence $U_d:$ 
\begin{align}
    U_d &= R_z(\phi_0) \left(\prod_{j=1}^{d} {\rm JC}(\theta, \varphi) R_z(\phi_j)\right).
\end{align}
By the structure of Eq.~\eqref{eq:JC-as-QSP-block}, the above sequence performs QSP within each $2\times 2$ subspace spanned by $\{ \ket{e, n}, \ket{g, n+1}\}$, such that the QSP polynomials obtained are 
\begin{align}
    \label{eq:JC_QSP_poly-block}
    \begin{bmatrix}
        P(x_{n,n+1}) & -i\sqrt{1-x_{n, n+1}^2}Q(x_{n,n+1})
        \\
        -i\sqrt{1-x_{n, n+1}^2}Q^{*}(x_{n,n+1}) & P^{*}(x_{n,n+1})
    \end{bmatrix}
\end{align}
where the polynomials are of the form $P(x_{n,n+1}) = \sum_{m=0}^{d}p_{m}x^m_{n,n+1}$ and $Q(x_{n,n+1}) = \sum_{m=0}^{d-1}q_{m}x^m_{n,n+1}$ for coefficients $p_{m}, q_m \in \mathbb{C}$. 
By choosing $P$ and $Q$ such that they attain the desired values at the discrete signal points $x_{n,n+1}$, we can realize a variety of Fock-state-selective operations in cavity QED, as we discuss in the following subsections.

\subsection{Selective Manipulation of Single Fock State}
\label{subsec:qsp-ctrl:selphtn}

As we mentioned above, the Jaynes-Cummings interaction, combined with programmable single-qubit rotation gates in the form of a QSP construction provides a mechanism for selectively adding or subtracting a single photon from a particular Fock level.

To realize a selective increment operation that maps the Fock state $\ket{r}$ to $\ket{r+1}$, we require a polynomial satisfying
\begin{align}
    P(x_{r,r+1}) \approx 0,~~~~~~
   \sqrt{1- x_{r,r+1}^2}   Q(x_{r,r+1}) \approx -i
\end{align}
at the target Fock level $r$.
Under this condition, the corresponding $2\times2$ block associated with the subspace spanned by $\{\ket{e,r},\ket{g,r+1}\}$, as shown in Eqs.~\eqref{eq:JC-sub-block} and~\eqref{eq:JC_QSP_poly-block} becomes proportional $-iY$.
This implements complete transfer of excitation from the qubit to the oscillator, thereby mapping the Fock state $\ket{r}$ to $\ket{r+1}$.

The action of the selected $2\times 2$ block depends on the initial state of the qubit. In the basis $\{\ket{e,r},\ket{g,r+1}\}$, one column of the block determines how the state evolves when the qubit begins in $\ket{e}$, while the other column is relevant when the qubit begins in $\ket{g}$. Consequently, the condition for selective photon addition or subtraction depends on which direction of excitation transfer is desired.

For a selective increment operation, the qubit is first prepared in the excited state so that the excitation is transferred from the qubit to the oscillator. In this case, the block should act as a complete transfer from $\ket{e,r}$ to $\ket{g,r+1}$, which requires the relevant off-diagonal entry to satisfy
\begin{align}
    \sqrt{1-x_{r,r+1}^2}\,Q(x_{r,r+1}) \approx -i.
\end{align}
Conversely, for a selective decrement operation, the qubit begins in the ground state so that it absorbs an excitation from the oscillator. Then the transfer is from $\ket{g,r+1}$ to $\ket{e,r}$, and the corresponding condition is
\begin{align}
    \sqrt{1-x_{r,r+1}^2}\,Q^*(x_{r,r+1}) \approx -i.
\end{align}
Thus, the same block structure can realize either photon addition or photon subtraction, with the appropriate condition determined by the initial qubit state and the desired direction of excitation transfer.

\begin{figure}[ht]
    \centering
    \includegraphics[scale=0.87]{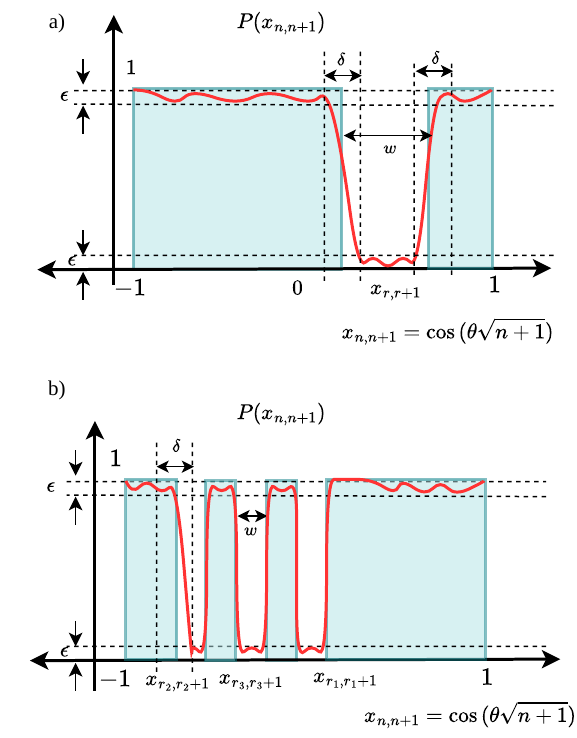}
    \caption{ Polynomial $P(x_{n,n+1})$ for which the incrementor increments photon number for a certain Fock number $r$ with error tolerance $\epsilon$ and resolution window $w$ with $\delta$ transition region is depicted in a). The polynomial for selective Fock state increment over multiple selected Fock levels is shown in b).}
    \label{fig:cavityQED-notch}
\end{figure}

For all other Fock levels to remain unchanged during this process, 
\begin{align}
    P(x_{n,n+1}) \approx 1, ~~~
      \sqrt{1- x_{n,n+1}^2} Q(x_{n,n+1}) \approx 0,  ~~\forall n \ne r , 
\end{align}
so that its action on the other $2\times 2$ blocks are equivalent to the identity. Thus, the desired polynomial in this case is analogous to a notch filter that is close to $0$ for the Fock level $r$ that we wish to manipulate, and is 1 for all other Fock levels:
\begin{align}
    P(x_{n,n+1}) &\approx \begin{cases}
        0,  ~n = r
        \\
        1,  ~n \ne r ,  
    \end{cases}
    \\
      \sqrt{1- x_{n,n+1}^2} Q(x_{n,n+1}) &\approx \label{eq:incr_q_poly}
     \begin{cases}
        \pm i, ~ n = r
        \\
        0, ~ n \ne r. 
    \end{cases}
\end{align}
We depict such a filter function $P(x_{n,n+1})$ in Fig.~\ref{fig:cavityQED-notch}.
With this transformation implemented, the ideal unitary operator $I_{+, \{r,r+1\}}$ takes the form, 
\begin{align*}
    I_{+, \{r,r+1\}} \approx
    \begin{bmatrix}
        1 & \mathbf{0}_{1 \times 2}  & \mathbf{0}_{1 \times 2} & \mathbf{0}_{1 \times 2} & \cdots
        \\
        \mathbf{0}_{2 \times 1} & I|_{(0, 1)}
      & \mathbf{0}_{2 \times 2} & \mathbf{0}_{2 \times 2} & \cdots
        \\
        \vdots & \vdots & \ddots & \vdots & \cdots
        \\
        \mathbf{0}_{2 \times 1} & \mathbf{0}_{2 \times 1} & \mathbf{0}_{2 \times 2} &  - iY|_{(r, r+1)}  & \cdots
         \\
        \vdots & \vdots & \vdots & \vdots & \ddots
    \end{bmatrix} . 
\end{align*}
We refer to the above operator as an increment operator when the desired block realized is $-iY$, which maps the Fock state $\ket{r}$ to $\ket{r+1}$, while the above operator acts as a decrement operator, denoted by $I_{-, \{r,r+1\}}$ when the block is $iY$.

In practice, the QSP polynomial $P(x_{n,n+1})$ can only approximate the desired notch filter up to some error tolerance $\epsilon$ (as depicted in Fig.~\ref{fig:cavityQED-notch}). For this, we require,
\begin{align}
    |P(x_{n,n+1})| &\le \epsilon, \quad n = r, \\
    |1 - P(x_{n,n+1})| &\le \epsilon, \quad n \neq r.
\end{align}
That is, the polynomial should suppress the selected Fock transition indexed by $r$, while remaining close to unity on all other transitions.

To construct such a polynomial, we approximate the ideal notch filter using a combination of top-hat functions $F_{\rm{Hat}}(x)$, following the approach introduced in Sec.~\ref{subsec:Linearize-error-scaling}. As in that construction, we approximate each top-hat function by a polynomial with a transition region of width $\delta$. 
In the present setting, the relevant resolution scale is set by the separation between the target value $\cos(\theta\sqrt{r+1})$ and the neighboring values $\cos(\theta\sqrt{r'+1})$ for some Fock level $r'$. We denote by $\Delta$ the minimum such separation:
\begin{align}
    \Delta = \min_{r \neq r'} \left| \cos(\theta\sqrt{r+1}) - \cos(\theta\sqrt{r'+1}) \right|.
\end{align}
This quantity determines the narrowest notch that can still resolve the target cosine value from its nearest neighbors. To avoid overlap between adjacent transition regions, one should choose $\delta < \Delta$; in the limiting case, we may take $\delta = \mathcal{O}(\Delta)$.
Then, the degree $d$ of the polynomial required to approximate the target notch filter with overall error $\epsilon$ scales as
\begin{align} \label{eq:single-incr-complexity}
    d = \mathcal{O}\left(\frac{1}{\Delta}\log\frac{1}{\epsilon}\right).
\end{align}

Lastly, we note that this construction can be straightforwardly adapted to realize a selective decrement operator that maps $\ket{r+1}$ to $\ket{r}$, which we denote by $I_{-,\{ r, r+1\}}$. Instead of preparing the qubit in the excited state so that it transfers an excitation to the oscillator, the qubit can be left in the ground state so that it can absorb an excitation from the targeted Fock component. In both cases, the selectivity is enforced by the same QSP-based filtering mechanism, with the only difference being the direction of excitation transfer.

\subsection{Selective Manipulation of Multiple Fock States }
\label{subsec:qsp-ctrl:incr}

In this subsection, we generalize the construction to simultaneously manipulate multiple Fock states. Our aim is to develop an operator that acts selectively on chosen Fock components of an oscillator state while leaving the remaining components unchanged. Specifically, for an oscillator state of the form
\begin{align}
    \ket{\psi}_{\mathrm{osc}} = \sum_{i}c_i \ket{r_i},
\end{align}
where $\{i\}$ indexes the populated Fock levels, and
we seek to selectively increment or decrement a chosen subset of Fock levels $\{r_j\}$ for some indices $j$, without affecting the other populated Fock states. As in the problem statement at the beginning of this section, we again assume that Fock levels $|r_j \pm 1\rangle$ are unpopulated to avoid collisions.
For example, we could consider the state
\begin{align}
    \ket{\psi}_{\mathrm{osc}} &=
   c_0\ket{r_0}+  c_1\ket{r_1} + c_2\ket{r_2}
    + c_3\ket{r_3} + c_4\ket{r_4},
\end{align}
and seek to manipulate it by incrementing only the selected Fock components $\ket{r_1}$, $\ket{r_2}$, and $\ket{r_3}$. The target oscillator state would then be
\begin{align}
    \ket{\psi}'_{\mathrm{osc}} &= \nonumber
   c_0\ket{r_0} + c_1\ket{r_1+1} + c_2\ket{r_2+1}
   \\& + c_3\ket{r_3+1} + c_4\ket{r_4}.
\end{align}
We present two ways to achieve this in the following. 
At a high level, such selective manipulation of multiple Fock states $\{r_j\}$ can be achieved by composing several such selective photon-number operators. 
Operationally, utilizing the construction presented in Sec.~\ref{subsec:qsp-ctrl:selphtn} multiple times, we can implement this as
\begin{align}
   I_{\text{overall}} = (\sigma_x \otimes I_{\text{osc}}) \prod_{\{r_j\}} \left[I_{+, \{r_j,r_j+1\}} (\sigma_x \otimes I_{\text{osc}})\right] , 
\end{align}
where the term $\sigma_x \otimes I_{\text{osc}}$ is included for pumping additional photons into the qubit-oscillator system. Explicitly, we take the qubit to initially be in ground state $\ket{g}$, such that applying $\sigma_x \otimes I_{\text{osc}}$ excites it to $\ket{e}$ and then application of the incrementor operator transfers the excitation to the oscillator. 

Assuming that we intend to selectively manipulate a total of $N$ Fock states with $\Delta$ being the minimum separation needed to resolve the relevant cosine values, then the above construction requires a number of queries to the Jaynes-Cummings interaction that scales as
\begin{align}
 \mathcal{O}\Big( \frac{N}{\Delta}\log{\frac{1}{\epsilon}} \Big) .
    \label{eq:degree-bound-naive}
\end{align}
This is equivalently the sum of the polynomial degrees implemented by QSP, and scales linearly with the number of Fock levels to be manipulated. 

\textbf{An exponentially better construction.} While the above approach performs selective manipulation of multiple Fock states sequentially, a more efficient construction can be developed by implementing a QSP polynomial that performs selective manipulation all at once. Here we show how to achieve this. 

Let us focus on performing selective incrementation of Fock levels $\{ r_j\}$ for simplicity. An individual QSP polynomial can achieve this for all $\{ r_j\}$ if it obeys
\begin{align}
    P(x_{n,n+1}) \approx \begin{cases}
        0, \quad \forall n \in \{r_j\}
        \\
        1, \quad \text{otherwise.}
    \end{cases}
\end{align}
The complementary polynomial $Q(x_{n,n+1})$ follows a form similar to Eq.~\eqref{eq:incr_q_poly}.
This corresponds to an overall operator $I_{\rm{overall}}$ that takes the form,
\begin{align*}
    I_{\rm{overall}} 
    \approx
    \begin{bmatrix}
        1 & \mathbf{0}_{1 \times 2}  & \mathbf{0}_{1 \times 2} & \mathbf{0}_{1 \times 2} & \cdots 
        \\
        \mathbf{0}_{2 \times 1} & \pm iY|_{(0, 1)}
      & \mathbf{0}_{2 \times 2} & \mathbf{0}_{2 \times 2} & \cdots
        \\
        \vdots & \vdots  & \ddots & \cdots & \cdots 
        \\
        \mathbf{0}_{2 \times 1} & \mathbf{0}_{2 \times 2} &  \mathbf{0}_{2 \times 2} &  \pm iY|_{(r_j, r_j+1)}  & \cdots 
        \\
      \mathbf{0}_{2 \times 1} & \mathbf{0}_{2 \times 1}
      &\mathbf{0}_{2 \times 1} & \mathbf{0}_{2 \times 2} & \ddots 
    \end{bmatrix} . 
\end{align*}

Alternatively, we note that an analogous construction holds for selective decrementing of multiple Fock levels. Here, if a particular Fock state in the superposition is to be decremented, then the qubit can be left in the ground state $\ket{g}$ to allow it to receive the photon removed from the targeted Fock state. 
It is important to note that the Jaynes-Cummings interaction conserves the total excitation number of the joint qubit-oscillator system. Consequently, photon-number increment or decrement operations on the oscillator are realized through excitation exchange with the qubit.

The polynomial degree required to realize $P(x_{n,n+1})$ depends on both the target error tolerance $\epsilon$ and the minimum resolvable separation $w$ between the cosine values associated with different Fock levels. For a protocol that selectively manipulates a total of $N$ Fock states, the required degree scales as
\begin{align}
    d = \mathcal{O}\!\left(\frac{1}{\Delta}\log\frac{N}{\epsilon}\right).
\end{align}
Here as stated previously, $\Delta$ denotes the minimum separation needed to resolve the relevant cosine values, and $N$ is the number of Fock states to be manipulated selectively. This scaling follows from Eq.~\eqref{eq:single-incr-complexity} by selecting each polynomial to suffer individual error $\epsilon/N$ such that the total error is at most $\epsilon$.  This bound scales logarithmically with the number of Fock levels to be manipulated, which is notably an exponential improvement over the complexity of the sequential approach in Eq. \eqref{eq:degree-bound-naive}.

\section{Discussion and Conclusions}
\label{sec:conclusion}

The scaling and growth of quantum computers in the near future inherently depends on advances in quantum control techniques. An improved understanding of the control methods governing qubit-oscillator dynamics is, therefore of fundamental importance to ensure scalability of the present-day hardware. 
In this work, we take inspiration from the rich field of quantum algorithms to propose QSP-control as an analytic framework for quantum control of qubit-oscillator systems with specific applications to the building-blocks of superconducting quantum devices, namely circuit-QED and cavity-QED. 
We show how the problem of linearizing the effects of Kerr interactions in a qubit dispersively coupled to a microwave cavity in circuit-QED systems can be recast as a polynomial approximation problem and effectively solved with a bounded error. This could potentially improve the performance of the transmon qubit without any external hardware engineering.

In the context of cavity-QED systems, we mapped the Jaynes-Cummings gate to a QSP construction to develop unitary operators capable of selectively addressing individual Fock states of an oscillator, and incrementing or decrementing them to the target levels while keeping other Fock states unchanged. Such operators can be utilized to prepare specialized superposition states used in encoding quantum information into oscillators, in error-correcting codes such as bosonic binomial codes, and in constructing qudits. 

While the analytic guarantees of QSP allows the design of precise transformations that address these problems, the corresponding QSP polynomials often have high degree, usually with circuit depths of the order of a few hundreds to obtain the desired error tolerance. This limits its applicability in systems where fast control and appropriate decoherence time (qubit $T_1,$ and $ T_2$) are not readily available. 
Another limitation of the present framework is that its performance depends on the spectral resolvability of the relevant Fock-state transitions and on the polynomial degree required by the QSP construction. As the number of selectively controlled Fock levels increases, these requirements may lead to longer sequences and tighter control constraints.

Possible future directions for this work include exploring other methods of coupling qubits to oscillators in a QSP formalism, beyond the dispersive and Jaynes-Cummings couplings studied in this work. A key objective in this pursuit will be identifying other operations that admit structures analogous to the SU(2) rotations or block encodings used in QSP. Further, it is yet to be explored how we can extend this work to multi-qubit coupled multi-mode systems, where multivariate generalizations of QSP~\cite{laneve2024quantumsignalprocessingsun, Lu2026quantumsignal} can be utilized to realize appropriate polynomials. 
Beyond the applications considered here, the same framework may prove useful for preparing other non-classical oscillator states. One promising direction is the preparation of NOON states \cite{PhysRevLett.121.160502, wfj8-tgjz}, which are relevant to quantum metrology \cite{Mitchell2004,Velasco2025} and quantum communication \cite{Grun2022}.
In aggregate, the developments of this work open new avenues for tackling quantum control problems through the analytic lens of quantum algorithms and polynomial approximation theory. 
\begin{acknowledgments}
    This work is supported by the U.S. Department of Energy, Office of Science, Advanced Scientific Computing Research, under contract number DE-SC0025384. This research was supported by PNNL’s Quantum Algorithms and Architecture for Domain Science (QuAADS) Laboratory Directed Research and Development (LDRD) Initiative. This
    material is based upon work supported by the U.S. Department of Energy, Office of Science, National Quantum Information Science Research Centers, Quantum Science Center (QSC). The Pacific Northwest National Laboratory is operated by Battelle for the U.S. Department of Energy under Contract DE-AC05-76RL01830. 
\end{acknowledgments}


\bibliography{prx}

\end{document}